\documentstyle[12pt,fleqn]{article}
\begin{document}
\hfill{\large RPI-98-N115}
 
\hfill{\large FTUV 98/19}

\hfill{\large IFIC 98/19}
\vspace{.2cm}

\begin{center}
{\large{\bf Inclusive Neutrino Scattering in $^{12}C$ : 
Implications for $\nu_{\mu}$ to $\nu_e$  Oscillations}}
\end{center}

\vspace{1cm}

\begin{center}
{\large S. K. Singh$^{a, b}$, Nimai C. Mukhopadhyay$^{a, c}$
and E. Oset$^a$}
\end{center}

\vspace{0.5cm}

$^a$ {\small{\it Departamento de F\'{\i}sica Te\'orica and IFIC,
Centro Mixto Universidad de Valencia-CSIC,
46100 Burjassot (Valencia), Spain}}

$^b$ {\small{\it Physics Department, Aligarh Muslim University, Aligarh, India
202002}}

$^c$ {\small{\it Department of Physics, Applied Physics and Astronomy, 
Rensselaer Polytechnic Institute, Troy, New York 12180-3590}}

\begin{abstract}
We study inclusive $\nu_e$ and $\nu_{\mu}$ cross sections
in $^{12}C$ in a theory that takes into account significant nuclear 
renormalization of strengths. Our calculation is in excellent agreement 
with the measured inclusive muon capture rate and 
the flux-averaged $\nu_e$ cross
section, but overestimates the flux-averaged $\nu_{\mu}$ inclusive cross
section. These reactions are  of crucial relevance to the issue of
 possible 
$\nu_{\mu}$ to $\nu_e$ oscillations.We also calculate the flux-averaged
cross sections  in $^{13}C$ and $^{27}Al$,  which are found to be
consistent with the available experimental result.
\end{abstract}

\vspace{0.5cm}

\newpage
\begin{center}
{\bf I.INTRODUCTION}
\end{center}

Extracting an interesting physics signal of some rare events from an experiment
crucially depends on our ability to understand the physics background in that
setting. A good example of this is the search for neutrino oscillations, 
and hence physics beyond the standard model, in the Liquid Scintillator
Neutrino Detector (LSND) 
related experiments, with profound implications
for particle physics, nuclear physics and astrophysics. 
The LSND group have advanced
 evidence for the $\nu_{\mu} \rightarrow \nu_e$ oscillations in 
a recent experiment with neutrinos
from pion decays in flight \cite{1}. Our trusts in this claim  depend crucially
upon the performance of these experiments in
benchmark reactions on nucleons and nuclei initiated by neutrinos.
Some of these benchmark reactions are 
exclusive ones on proton, $^{12}C$ and $^{16}O$ targets, while others
are inclusive processes in complex nuclei. 
Given separate claims of agreements with experiment and lack of it
in recent theoretical calculations \cite{2,3,4,4a,4b} of inclusive neutrino 
reactions in $^{12}C$, this  is of  topical interest and deserves
a careful examination.

In this paper, we are going to discuss inclusive neutrino reactions in 
nuclear targets
$^{12}C$, $^{13}C$ and $^{27}Al$,
concentrating on the  theory of the reactions \cite{5}
$$
^{12}C + \nu_e \rightarrow  X + e^-,
\eqno{(1)}
$$

$$
^{12}C + \nu_{\mu} \rightarrow  X + \mu^-,
\eqno{(2)}
$$
\noindent
where the produced $X$ is not observed. Thus, generically, we can 
indicate the reactions (1,2) as $^{12}C (\nu_l, l^-) X $ where $X$
is the unspecified nuclear state and $l$ is the charged lepton. At low
energies, the reaction (1) has been measured by 
the E225 \cite{6} and 
LSND \cite{7,8} experiments at the Los Alamos Meson
Physics Facility (LAMPF) and the experiments by KARMEN
\cite{9,10}
collaboration at the ISIS facility,
with $\nu_e$beam from muon decays at rest (DAR), given by the
Michel spectrum. The flux-averaged cross sections from the various experiments
are given in Table I. 
Particularly interesting to us are the most
recent values of the cross section for the reaction (1) 

$$
\sigma (\nu_e) = (14.8 \pm 0.7 \pm 1.1) \times 10^{-42} \, cm^2,
\eqno{(3a)}
$$

\noindent
from the LSND experiment \cite{8} and

$$
\sigma (\nu_e) = (15.2 \pm 1.0 \pm 1.3) \times 10^{-42} \, cm^2,
\eqno{(3b)}
$$
\noindent
from the KARMEN collaboration \cite{10}. In Eq.(3), the inclusive 
cross sections
are obtained by adding the cross sections leading to ground state and 
excited states, with their errors added in quadrature. Within their errors,
these experimental results overlap. The reaction (2), on the other hand, has
been measured over the last five years
at LAMPF \cite{11,12,13}
using $\nu_{\mu}$ beam from pion decays in flight (DIF). The earlier
E764 
experiment of Koetke {\it et al.} \cite{11} used a  neutrino beam of slightly
higher energy than used in later experiments and 
gave a 
cross section for the reaction (2) too large when compared to the theoretical
predictions \cite {2,3,4,4a,4b}.
This reaction was further studied
by Albert {\it et al.} \cite{12} with more massive 
detectors and larger exposure
than used in \cite{11},  using a  beam of 
neutrinos with energy
around 180 MeV.
These authors 
obtained considerably smaller cross section than 
that by Koetke {\it et al.} The recent
studies by the LSND collaboration 
\cite{13} have improved 
the experimental situation, providing the 
latest measured value for this cross section. We give, in Table I, the 
results of all of these experiments,  but use in our
theoretical discussion the latest 
result, reported by the LSND collaboration \cite{13}
$$
\sigma (\nu_{\mu})= (11.2 \pm  0.3 \pm 1.8) \times 10^{- 40} \, cm^2.
\eqno{(4)}
$$
We also calculate the flux-averaged neutrino cross-sections 
in $^{13}C$ and $^{27}Al$
as benchmarks for our theory. These are found to be reasonably consistent
with the available experimental results  \cite{6}.

Current generation of calculations of the above reactions can be grouped
into two classes, depending on their predictions for the flux-averaged
cross sections for the process (2): those that {\it substantially exceed} this
observed cross section \cite{2,3,5} and others that {\it 
find agreement}
\cite{4,4b}. The deficit of the flux-averaged cross sections may be: (1) a
manifestation of theoretical problem, of not being able to do a correct
enough nuclear calculation; (2) an experimental problem of not doing a precise
and reliable
enough experiment. Since both theoretical calculations and experimental
analysis are involved in the determination of the excess events in the 
$\nu_e$ channel in the LSND neutrino oscillation experiment \cite{1}, it
is important to have a clear understanding of the nuclear physics
related uncertainties in this reaction.
The purpose of this paper is to narrow down the options by examining 
the first point very critically from our point of view \cite{5}.

A reaction which can be regarded as a benchmark in the context of the processes
(1) and (2) in general, and (2) in particular, is the nuclear capture of 
muons (NMC) from the atomic $1 S$ state by the charged weak current
 \cite{14,15}:

$$
^{12}C + \mu^- (1S) \rightarrow  X + \nu_{\mu}.
\eqno{(5)}
$$

\noindent
This process  serves
as an excellent check, in the low and intermediate energy transfer region, in
our ability to control the theoretical uncertainties \cite{15}. 
The inclusive NMC rate $\Lambda_c$ for the process (5) is very accurately
known. Taking the world average of the best experimental determinations
of the inclusive muon capture rate $\Lambda_c$,\cite{16,17}with their errors 
added in quadrature, we obtain\cite{15},

$$
\Lambda_c = (3.80 \pm 0.10)\times 10^4 s^{- 1}.
\eqno{(6)}
$$
\noindent
Thus, we have here a weak reaction rate, which is closely related to the
processes (1) and (2) and is known at an accuracy of about $2.5 \, \%$,
posing a tremendous challenge to the theoretical approaches to weak nuclear
reactions in nuclei.

Various theoretical approaches
to calculate reactions (1) and (2) \cite{2,3,4,4a},
when applied
to the inclusive muon capture, reproduce the NMC rate $\Lambda_c$ quoted
in Eq.(6) rather well, within the limits of their theoretical accuracy
\cite{4,16,19}.
In the case of inclusive neutrino reactions, however,
the situation is different.
The calculations of Kolbe {\it et al.} reproduce the $(\nu_e,e^-)$ rather well,
but overestimate the $(\nu_{\mu},\mu^-)$ by about $50\%$.
 Auerbach {\it et al.},
 on the other hand, 
use 
the set of parameters  for their model,
which explain the inclusive 
muon capture rate. They predict cross sections for 
the neutrino reactions, which 
are
$20\%$ and $15\%$ larger than the experimental values for the reactions (1) and
(2) respectively, quoted in Eqs.(3) and (4). These overestimates are 
from the maximum experimental values  allowed within the 
quoted errors. 
 Therefore, they represent {\it a real discrepancy} with the experiment.
It is possible to explain the observed neutrino cross sections
in the calculations of Auerbach {\it et al.} with another version of their 
residual nuclear forces by varying the model parameters, but this version
predicts a NMC capture rate of $ 3.09\times 10^4s^{-1}$, which is rather
 small compared to the value quoted in Eq.(6).

Finally, to complete this survey,
we would like to mention that a Fermi gas model calculation with a
Fermi momentum $k_F = 225 \, MeV$ gives a much higher value of 
$24.1 \times 10^{- 40} \, cm^2$ for the flux-averaged 
$(\nu_{\mu}, \mu^-)$ cross
section. This is reduced to $22.7 \times 10^{- 40} \, cm^2$, when the effects
of meson exchange currents are taken into account \cite{3}. In another
 calculation , the method of so-called ``elementary particle model'', extended 
to inclusive reactions, has been used, to obtain a cross section of
$13.1\times 10^{-40} \, cm^2$ \cite{4b}, which is 
in good agreement with the experimental
value.
However, the extension of the elementary particle model to the
inclusive reactions makes use of several assumptions, which have not 
been tested in the energy region of the LSND experiments. Here this
 method is 
expected to underestimate the inclusive cross
 sections \cite{4a,12,20}.

This survey brings us to our calculation, which we briefly describe in section
II and discuss the results and conclusions in section III.

\begin{center} 
 {\bf II. FORMALISM}
\end{center}

The matrix element for the neutrino nucleon reaction for a neutrino of flavor
$l, (l = e, \mu)$, i e.
     
$$ 
\nu_l (k) + n (p) \rightarrow l^- (k') + p (p'),
\eqno{(7)}
$$

\noindent
is given by

$$
T = \frac{G cos \theta}{\sqrt{2}} \bar{u} (k') \gamma^{\mu} (1 - \gamma_5)
u (k) J_{\mu},
\eqno{(8)}
$$

\noindent
where

$$
J_{\mu} = \bar{u} (p') [F_1^V (q^2) \gamma_{\mu} + F_2^V (q^2) i 
\sigma_{\mu \nu} \frac{q^{\nu}}{2 M} + F^{V}_A (q^2) \gamma_{\mu} \gamma_5 +
F^V_P (q^2) q_{\mu} \gamma_5] u (p).
\eqno{(9)}
$$

\noindent
In Eq. (9), $q_{\mu} = k_{\mu} - k'_{\mu}$, is the four momentum transfer,
$F^V_1, F^V_2, F^V_A$ and $F^V_p$ are the known weak nucleon form factors.
The double differential cross section $\sigma_0 (q^2, k')$, is then given by

$$
\sigma_0 (q^2, k') = \frac{k'}{4 \pi E E'} \, \frac{M^2}{E_n E_p} \,
\bar{\Sigma} \Sigma |T|^2 \delta (E - E' + E_n - E_p),
\eqno{(10)}
$$
\noindent
where $\bar{\Sigma} \Sigma |T|^2 $ 
represents the sum and the average respectively
over the final and the initial spins of the leptons and the nucleons and
is evaluated exactly, using the matrix element $T$ defined in Eq.(8). 
Its analytic
expression is given in ref. \cite {5}. 
In a nucleus, the neutrino scatters from a
neutron moving in the finite nucleus of neutron density 
$\rho_n (\vec{r})$, with a
local neutron occupation number $n_n (\vec{p},\vec{r})$. 
Then the cross section in
the local density approximation and in the free nucleon picture is given by

$$
\sigma (q^2, k') = 2 \int d \vec{r} \frac{d \vec{p}}{(2 \pi)^3} n_n
(\vec{p}, \vec{r}) \sigma_0 (q^2, k'),
\eqno{(11)}
$$
where the neutron energy $E_n$ and the proton energy $E_p$ 
in the expression of $\sigma_0(q^2, k')$, given in Eq.(10), are 
replaced by $E_n(p)$ and $E_p(p')$ respectively, $p$
and $p'$ being the momenta of the neutron and proton in the nucleus.
However, 
neutrons and protons are not free and 
their momenta are constrained
to satisfy the Pauli principle $i. e. \, p < p_{F n}$ and $p' > p_{F p}$, where
$p_{F n}$ and $p_{F p}$ are the local Fermi momenta, given by

$$
p_{F n} (r) = [3 \pi^2 \rho_n (r)]^{1/3} \; \hbox{and} \;
p_{F p} (r) = [3 \pi^2 \rho_n (r)]^{1/3}. 
\eqno{(12)}
$$
Moreover, in the finite nucleus, there is a threshold energy
for the reaction to proceed, also called the Q-Value,
and this should be taken into account. Finally, the charged lepton produced in
reactions (1) and (2) moves in the nucleus and its energy is modified by  the
Coulomb interaction, which should be accounted for. In our approach, these
effects are incorporated by modifiying the argument of the $\delta$ function 
in Eq. (10), from $E - E' + E_n - E_p$ to $E - E' - V_c (\vec{r}) +
E_n (p) - E_p (p')$, and replacing the factor
$\int \frac{d \vec{p}}{(2 \pi)^3} n_n (\vec{p}, \vec{r}) \frac{M^2}{E_n E_p} 
\delta (E - E' + E_n - E_p)$ occurring
in Eq. (11) by $- \frac{1}{\pi} Im U (q_0, \vec{q})$, where

$$
q_0 = E - E' - V_c - Q + Q'.
\eqno{(13)}
$$
\noindent
In Eq.(13),   $ V_c(\vec{r})$ is the Coulomb energy of the lepton
and  $Q' = E_{F p} - E_{F n}$,  is introduced to take into
account the unequal Fermi sea in the case of $N \neq Z$ nuclei.

\noindent
$U(q_0,\vec{q}$, is the Lindhard function given by

$$
U (q_0,\vec{q}) = \int \, \frac{d^3 p}{(2 \pi)^3} \,
\frac{n_n (\vec{p}) [1 - n_p (\vec{p} + \vec{q})]}
{q_0 + E_n (\vec{p}) - E_p (\vec{q} + \vec{p}) + i \epsilon} \,
\frac{M^2}{E_n E_p}.
\eqno{(14)}
$$
With these modifications, the total cross section 
$\sigma (E_{\nu})$ is given as \cite{5}:

              $$
\sigma (E_{\nu}) = - \frac{4}{\pi} \,
\int^{\infty}_0 r^2 d r \, \int^{p_l^{max}}_{p^{min}_l} \,
k'^2 dk'  \, \int^1_{- 1} d (\cos \theta) \, 
\frac{1}{E_{\nu} E_l} 
$$

$$
\times \bar{\Sigma} \Sigma |T|^2 I m U
[E_{\nu} - E_l - Q + Q' - V_c (r) , \vec{q}] \Theta
[E_l + V_c (r) - m_l].
\eqno{(15)}
$$

\noindent
The kinematic limits $p_l^{max, min}$ for the lepton momentum $k'$
 are easily computed in our
special case\cite{5}. For the numerical integrations, we use 
Gaussian quadrature with high
enough accuracy for our purpose. The radial integration in Eq. (15) is
performed up to a radius $R = c_1 + 5 fm$, where $c_1$, is the radius 
parameter in the two-parameter  harmonic oscillator and Fermi density
distributions  used for the nuclei, considered in section III.

The renormalization of weak currents in the nuclear medium is taken into 
account by calculating the effect of propagation of the particle-hole (ph)
excitations in the nuclear medium on various terms occurring in
$\bar{\Sigma} \Sigma |T|^2$. The ph response is then 
replaced by a Random Phase Approximation (RPA)
response accounting for the ph and the $\Delta h$ 
components, which interact through
an effective spin-isospin nuclear interaction described by the Landau-Migdal 
potential. The details of this renormalization procedure as well as those 
of Eq. (15) are
given in \cite {5}. However, we have made here the  following
 improvements, which considerably reduce the theoretical uncertainties 
in our calculations 
from the previous
versions of our model: (1) Our new Lindhard
function makes use of a strategy \cite{21} that avoids the pathologies of
the ordinary \cite{5} Lindhard functions in the limit of 
$q_0, \vec{q}$ both going to zero,
$q_{\mu}$ being the momentum transferred to the nucleus 
in the processes of interest. (2) The nuclear response function is renormalized
by the ph and the $\Delta h$
correlations in nuclei \cite{5}, effects of which are quite large for 
low and intermediate
energy neutrino scattering. The physics of this renormalization depends, 
among other things, on the Landau-Migdal spin-isospin parameter
$g'$ \cite{22}. This itself has an uncertainty
of $\pm 0.1$ around its preferred value of $0.7$ \cite{23}. We take into 
account the theoretical uncertainties of our estimates of $\sigma (\nu_e)$ 
and $\sigma (\nu_{\mu})$ due to this variation of $g'$. (3) Finally, 
the target
nucleus $^{12}C$ has intrinsic parametric uncertainties in the radial density
function. The effect due to this
uncertainty in our cross-section estimate is taken into account by repeating
our calculation in several radial parametric settings \cite{24}. Overall,
we achieve a theoretical accuracy around  $\pm 10 \%$ for 
$\sigma (\nu_e)$ and $\sigma (\nu_{\mu})$.
{\it This significant improvement in theoretical accuracy} is even better
in the case of the NMC rate for the inclusive process (5) \cite{16}. 
We estimate here an uncertainty of $\pm 2 \%$ due to nuclear radial effects 
and $\pm 5 \%$
due to the variation of the spin-isospin parameter $g'$. Treating these
two uncertainties independently and adding them in quadrature, we get a 
theoretical error of about $\pm 6 \%$ and obtain \cite{16}

$$
\Lambda_c = (3.60 \pm 0.22 ) \times 10^4 s^{-1}, 
\eqno{(16)}
$$

\noindent
{it in excellent agreement} with the precise experimental data (Eq.6).
The inclusive nuclear muon capture provides us with a critical benchmark,
an independent accurate check of our ability to
describe nuclear inclusive weak processes clearly related to the neutrino
scattering.

In summary, our method used in this paper is essentially an RPA approach
built up
from single particle states of an uncorrelated local Fermi sea. This method
is, in practice, found to be a very accurate tool, 
when the excitation energy is
sufficiently large such that relatively many states contribute to the 
process, in particular, if a large fraction of it comes from excitation 
to the continuum, as it is in the present case. The adaptation of this method 
to finite nuclei via the local density approximation has proved to be rather 
advantegeous to deal with inclusive reactions and has been successfully
applied to the photonuclear reactions\cite{25}, electron scattering\cite{26},
deep inelastic scattering \cite{27} and muon capture\cite{16,28}.

The  numerical evaluation of  the neutrino-nucleus reaction cross section 
is done using Eq.(15) and the rsults are presented in section III below.

\begin{center}
{\bf III. RESULTS AND CONCLUSIONS}
\end{center}

          In order to compare with the experimental results of KARMEN\cite{10}
and LSND \cite{6} collaborations, we compute the flux-averaged cross section,
$$
\bar{\sigma} = \frac{\int_0^{E_\nu^{max}} \, \sigma (E_\nu) \, \omega (E_\nu)
\, d E_\nu}{\int_{E_0}^{E_\nu^{max}}
 \omega \, (E_\nu) \, d E_\nu},
\eqno{(17)}
$$

\noindent
where the neutrino profile function $\omega (E_{\nu})$ is well-known
(i.e., the Michel spectrum in $E_{\nu_e}$ and the spectrum of $E_{\nu_{\mu}}$
provided by the LSND experiment). The lower limit $E_0$ in
Eq.(17) is taken to be zero for the $(\nu_{e}, e^-)$
reaction and $123.1$ MeV for the $(\nu_{\mu}, \mu^-)$ reaction \cite{11}.

For $^{12}C$, we present our  results in Tables II and III, 
and compare with experiments and other recent theoretical
works in Table IV. Here are the main points of our analysis. In
both Tables II and III, the rows 1 through 4 indicate four different choices of
the radial parameters for the nuclear density \cite{24}. The columns
represent different choices of the Landau-Migdal spin-isospin parameter
$g'$. In Table III, we have used  the $\nu_{\mu}$ 
spectrum reported
in the LSND papers \cite{1,11} for a direct comparison with the experiments.
The radial uncertainties are
typically about $2 \%$, while the $g'$ variation represents a $\pm 7 \, \%$
spread, for the $(\nu_e, e^-)$ case and $\pm 8 \, \%$ for the 
$(\nu_{\mu}, \mu^-)$ case
both around the central values corresponding to $g' = 0.7$. Thus,
the overall spread from the theoretical error, taking both of these effects
in quadrature, is $\pm 7.3 \%$ for the $(\nu_e, e^-)$ case and 
$\pm 8.2 \, \%$ for the $(\nu_{\mu}, \mu^-)$ case. In
Table IV, we compare the presently available theoretical results with 
the most recent experimental results for these reactions.

From Table IV, we can provide our best estimate of $\bar{\sigma} (\nu_e)$ and
$\bar{\sigma} (\nu_{\mu})$ as follows:

$$
\bar{\sigma} (\nu_e) = (15.48 \pm 1.13) \times 10^{- 42} cm^2,
\eqno{(18)}
$$

$$
\bar{\sigma} (\nu_{\mu}) = (16.65 \pm 1.37) \times 10^{- 40} cm^2.
\eqno{(19)}
$$

\noindent
We are in excellent agreement with the experimental 
value of $\sigma (\nu_e)$,
but {\it our lower limit $\sigma (\nu_{\mu})$ is $15 \, \%$ 
higher than the higher
limit of $\sigma (\nu_{\mu})$ measured by the LSND collaboration.}
 
    For $^{13}C$, we obtain a flux averaged cross section of 
$ 7.25\times 10^{-41}
 cm^2$ for Michel spectrum, using the density distribution parameters 
given in\cite{24}.This should be compared with the calculations of Arafune
{\it et al.} \cite{29}, who obtain a 
cross section of $ 9.58\times 10^{-41} cm^2$ for 
the transition to the ground state and first excited state of 
the final nucleus,
which together give $85 \, \%$ of the total coss section. This implies an
inclusive cross section of $ 11.3\times 10^{-41} cm^2$. There could be a
reduction of $(10-15) \, \%$,if the momentum dependence of the form 
factors are taken into account\cite{30}. This value seems to be in agreement 
with an unpublished result of Donnelly quoted by Krakauer {it et al.} \cite{6}.
The calcultions of Arafune {it et al.} \cite{29} do not take into account the
possible quenching of the weak interaction operators in nuclei, which is
studied by Fukugita {it et al.} \cite {31}.
In this paper, the quenching of the matrix elements of the Gamow-Teller (G-T) 
operators
is obtained in an effective operator approach, which takes into account
the effect of core polarization, isobar  and the meson exchange 
current processes. This leads to a $20\,\%$ reduction in the flux-averaged
cross section for the ground state transition, while the cross section to 
the first excited state is reduced by a factor 3. Assuming, as before, that
these two states together contribute $85\,\%$ of the total cross section, 
a flux-averaged cross section of $ 5.4\times 10^{-41} cm^2$ is inferred from
the calculations of Fukugita {\it et al.} \cite{31}.

  We find that 
our flux-averaged cross section for $^{13}C$, reported in this paper,
is $35\,\%$ smaller than 
that  obtained by Arafune {\it et al.} while it  is
$35\,\%$ larger than the results
of Fukugita {\it et al.} It will be interesting to test 
these predictions by measuring
this cross section in the low energy neutrino experiments with liquid
scintillators, where $^{13}C$ forms part of natural carbon. In the experiments 
of Krakauer {\it et al.} \cite{6}, it is reported that
$$
          \sigma_{av}=0.723 \bar{\sigma}(\nu_e^{13}C)+\bar{\sigma}
(\nu_e ^{27}Al) <  18.3\times 10^{-41} cm^2.
\eqno{(20)}
$$

In our approach, we calculate the flux averaged cross section in $^{27}Al$
to be $ 11.48\times 10^{-41} cm^2$ with $ g'=0.7$ and using one set of 
parameters
from ref. \cite{24}. This, along with the value obtained for the neutrino cross
section in $^{13}C$, gives a value of $16.5\times 10^{-41} cm^2$ for
 $\sigma_{av}$ 
in Eq.(20). We associate a theoretical uncertainty of $6\,\%$ due to 
g' and density variation of $^{27}Al$ on this average cross section.
This value  is consistent with the available experimental information 
on these reactions.
 
We would like to emphasize that the renormalization of nuclear strengths in
our model produces a reduction of about $40 \%$ in the $(\nu_e, e^-)$ cross
section to bring it in agreement with the experiment \cite{1}. Similarly
large reductions also occur in the $(\nu_{\mu}, \mu^-)$ case.

In summary, our calculations  show no discrepancy with the measured
 flux-averaged 
$\nu_e$ cross sections in $^{12}C$, like other authors. We also nicely
reproduce the measured inclusive muon capture rate, now known very accurately
\cite{17,18}. But {\it we see a  discrepancy},
at least by $15 \, \%$ in
 the flux-averaged $\nu_{\mu}$ cross section compared with the LSND
experiment \cite{11}, the theoretical prediction being higher than the
experiment. The discrepancies between the experimental and
various theoretical results for $(\nu_{\mu}, \mu^-)$ inclusive cross
sections in $^{12}C$ {\it should be taken  seriously}, in view of its
 implications in present
studies of neutrino oscillations. Our results for the case of $^{13}C$
and $^{27}Al$ are consistent with the only experimental limit available
at present. A high-quality experimental measurement 
of the inclusive cross section,
specially in $^{13}C$, will be very useful in understanding the quenching
of the G-T strengths in this nucleus in the light of the wide range of 
theoretical predictions for this reaction.

\vspace{1cm}

We thank H. C. Chiang, H. J. Kim, E. Kolbe, T. S. Kosmas and W. Louis for
various stimulating discussions and communications. Two of us (NCM and SKS)
have the great pleasure of thanking E. Oset for his warm hospitality at the
Universidad de Valencia. NCM is grateful for the financial support of 
``IBERDROLA de Ciencia y Tecnolog\'{\i}a" and acknowledges the partial
support of the U. S. Dept. of Energy. SKS thanks the 
Ministerio de Educaci\'on
y Cultura for  his sabbatical support. This work is also supported by CICYT
under contract AEN-96-1719.

\newpage
Table I: A summary of flux-averaged $\nu_e$ and $\nu_{\mu}$ cross sections
as obtained in various experiments done with $^{12}C$ target.
The unit for the $\nu_e$ cross section is 
$10^{-42} cm^2$ and for $\nu_{\mu}$ cross section is $10^{- 40} cm^2$.

\vspace{1cm}
 
\begin{tabular}{|c|c|c|c|}
\hline
  & LSND collab.  & LAMPF E225 & KARMEN collab.   \\
$\sigma (\nu_e)$ & 14.8$\pm$.7$\pm$1.1 \cite{7,8}  &14.1$\pm$2.3 \cite{6}
& 16.8$\pm$1.4$\pm$1.7 \cite{9}  \\
     &   &    & 15.2$\pm$1.0$\pm$1.3 \cite{10} \\
\hline
$\sigma (\nu_{\mu})$ & 8.3$\pm$.7$\pm$1.6 \cite{12} &  LAMPF E764 &  --- \\
      & 11.2$\pm$.3$\pm$1.8 \cite{13}  &159$\pm$26$\pm$37 \cite{11} &  \\
\hline
\end{tabular}
 
\vspace{1cm}

Table II: Flux-averaged $\nu_e$. Four radial parameter sets are chosen 
from the literature 
\cite{24}, with parameters  $c_1$ and $c_2$ in fms. The 
Landau-Migdal parameter $g'$ is 
taken as $g' = 0.7 \pm 0.1$. The unit for cross section is 
$10^{- 42} cm^2$.

\vspace{1cm}

\small{
\begin{tabular}{|c|c|c|c|c|}
\hline
 $c_1$ & $c_2$ & $g' = 0.6$ & $g' = 0.7$ & $g' =0.8$ \\
\hline
1.687 &1.067 & 16.87 & 15.69 & 14.72 \\
1.672 &1.150 & 16.65 & 15.49 & 14.53 \\
1.649 &1.247 & 16.30 & 15.15 & 14.20 \\
1.692 &1.082 & 16.99 & 15.80 & 14.52 \\
\hline
\end{tabular}}

\vspace{1cm}

Table III. Flux-averaged $\nu_{\mu}$ cross section. Radial parameter sets
and $g'$ values are as in Table II. The  $\nu_{\mu}$ flux is taken from
$S_1$ \cite{13}  The unit for cross section is $10^{-40} cm^2$.

\vspace{1cm}

\small{
\begin{tabular}{|c|c|c|c|c|}
\hline
 $c_1$ & $c_2$ & $g' = 0.6$ & $g' = 0.7$ & $g' =0.8$ \\
\hline
1.687 &1.067 & 18.35 & 16.82 & 15.61 \\
1.672 &1.150 & 18.19 & 16.67 & 15.45 \\
1.649 &1.247 & 17.91 & 16.38 & 15.17 \\
1.692 &1.082 & 18.45 & 16.92 & 15.70 \\
\hline
\end{tabular}}

\vspace{1cm}

Table IV. Summary of flux-averaged cross sections. Experimental results
are inferred by adding ground state and excited state contributions
for $(\nu_e, e^-)$.
Theoretical results are from Kolbe {\it et al.} \cite{2}, 
Auerbach {\it et al.} \cite{4},
Umino {\it et al.} \cite{3} and this work. The units are $10^{-42} cm^2$ for
$(\nu_e, e^-)$ and $10^{- 40} cm^2$ for $(\nu_{\mu}, \mu^-)$ cross sections.

\vspace{1cm}

\small{
\begin{tabular}{|c|c|c|c|c|c|}
\hline
  & Kolbe  & Auerbach  & Umino  & This work & Exp \\
  & {\it et al.} \cite{2} & {\it et al.} \cite{4} & {\it et al.} \cite{3}
 &  &  \\
$\sigma (\nu_e)$ &  &  &  & & 14.8 $\pm$ 1.0 $\pm$ 1.5 \cite{7,8} \\
  & 15.6 & 12.9 - 22.7 & - & 15.48 $\pm$ 1.13 
& 15.2 $\pm$ 1.4 $\pm$ 1.8 \cite{10}  \\
\hline
$\sigma (\nu_{\mu})$ & 19.3 - 20.3 & 13.5 - 15.2 & 22.7 - 24.1 & 
16.65 $\pm$ 1.37 & 11.20 $\pm$ .3 $\pm$ 1.8\cite{13} \\
\hline
\end{tabular}}

\end{document}